\documentclass[useAMS,usenatbib,usegraphicx]{mn2e}

\newcommand{\lta}{\mathrel{\hbox{\raise 0.6 ex \hbox{$<$}\kern
                   -1.8 ex\lower .5 ex\hbox{$\sim$}}}}                
\newcommand{\gta}{\mathrel{\hbox{\raise 0.6 ex \hbox{$>$}\kern
                   -1.7 ex\lower .5 ex\hbox{$\sim$}}}}
\newcommand{\Mloss}{\ensuremath{\,\mbox{\it M}_{\odot}{\rm yr}^{-1}}}                   
\DeclareMathAlphabet{\mathsc}{T1}{cmr}{m}{sc}
\newcounter{tmpioncntr}
\newcommand{\ion}[2]{\setcounter{tmpioncntr}{#2}%
\ensuremath{\mathrm{#1\;\mathsc{\roman{tmpioncntr}}}}}

\title[Abundance distributions in magnetic ApBp stars]
      {Abundance distributions over the surfaces of magnetic ApBp stars: theoretical predictions}
\author[G. Alecian]
{
G.~Alecian$^{1}$\thanks{E-mail:georges.alecian@obspm.fr}
\\
$^{1}$LUTH, Observatoire de Paris, CNRS, Universit{\'e} Paris Diderot, 
             5 Place Jules Janssen, F-92190 Meudon, France\\
}
\begin{document}

\date{Accepted 2015 September 22.  Received 2015 September 21; in original form 2015 April 29}

\pagerange{\pageref{firstpage}--\pageref{lastpage}} \pubyear{2015}

\maketitle

\label{firstpage}

\begin{abstract}
   {Recently published empirical abundance maps, obtained through (Zeeman)
   Doppler mapping (ZDM), do not \textbf{currently} agree with the abundance
   structures predicted by means of numerical models of atomic diffusion in
   magnetic atmospheres of ApBp stars. In a first step towards the resolution of
   these discrepancies, we present a state of the art grid of equilibrium
   abundance stratifications in the atmosphere of a magnetic Ap star with
   $T_{\rm eff} = 10000$\,K and $\log g = 4.0$. A description of the behaviour
   of 16 chemical elements including predictions concerning the over- and/or
   under-abundances over the stellar surface is followed by a discussion of the
   possible influence of presently neglected physical processes.
   } 
\end{abstract}


\section{Introduction}
\label{intro}

Magnetic ApBp stars are known to exhibit inhomogeneous distributions of chemical
elements over their surface. \citet{Babcock_PASP_1949} and later
\citet{BabcockBurd1952} observed a variable magnetic field in $\alpha{^2}$CVn
which could, as in the case of a number of other magnetic stars
\citep{Babcock1958}, be interpreted in terms of the so-called oblique rotator
model. Here the magnetic axis of the rotating star is inclined with respect to
the rotational axis \citep{Babcock1949}. \citet{DeutschDe1956} speculated that
the observed periodic spectral variations and radial velocity variations were in
some way related to the magnetic field and worked out a method to map the
abundances of various chemical elements. This early work made it clear that
there had to be some correlation between the magnetic fields of a number of Ap
stars and the abundance anomalies seen in the spectra.

Later on, several studies tried to map the abundances at the surface of these
magnetic Ap stars \citep[review by][]{KhokhlovaKh1994}. With spectra of better
S/N ratio and with the full Stokes $IQUV$ profiles available at high spectral
resolution, modern mapping methods allow the simultaneous determination of the
magnetic field geometry and the horizontal abundance distributions over the
stellar surface \citep[see for instance the study of $\alpha{^2}$CVn
by][]{SilvesterSiKoWa2014b}. Vertical stratifications of chemical elements in
the atmospheres of ApBp peculiar stars have also been detected in magnetic ApBp
stars (a recent review is given by \citealt{Ryabchikova2008}).
\citet{BaileyBaLaBa2014l} have published evidence for secular variations in the
abundances of ApBp stars, and it has also been shown that the pulsation
properties of rapidly oscillating Ap (roAp) stars are affected by abundance
spots caused by the magnetic field \citep{FreyhammerFrKuEletal2009}.

On the theoretical side, atomic diffusion driven by radiative accelerations has
been proposed by \citet{MichaudMi1970y} to explain abundance anomalies in Ap
star atmospheres. By incorporating the effects of a magnetic field on charged
ions of an element, \citet{VauclairVaHaPe1979} have managed to explain Si
abundance anomalies observed in a number of chemically peculiar magnetic stars
and have demonstrated the important role played by horizontal magnetic field
lines in the accumulation of silicon. Soon afterwards, \cite{AlecianAlVa1981} --
for silicon -- and \citet{MichaudMiChMe1981} -- for several metals -- have for
the first time quantified, by means of a theoretical approach, the effect of the
magnetic field geometry on atomic diffusion in the atmospheres of magnetic ApBp
stars.

More recently, \citet{AlecianAlSt2010} and \citet{StiftStAl2012} have modelled
the bi-dimensional distributions of various chemical elements in magnetic
atmospheres. Their computations were based on the numerical code
{\sc{CaratStrat}} which evaluates the radiative accelerations of a number of
metals by detailed opacity sampling, taking into account Zeeman desaturation,
and solving the polarised radiative transfer equation. Equilibrium abundance
stratifications due to atomic diffusion (they depend on magnetic field strength
and inclination) are derived in an iterative procedure. These bi-dimensional
results make it possible to guess the horizontal and vertical abundance
distributions to be found in stars with a given magnetic geometry, in the
simplest case assumed to correspond to an oblique rotator with a centred dipole.

These theoretical predictions can be (and sometimes have been) confronted with
empirical abundance maps derived with the help of (Zeeman) Doppler mapping (ZDM)
\citep[see][]{VogtVoPeHa1987}. As a rule, these comparisons have either been
inconclusive or have not resulted in agreement between predictions from theory
and the detailed surface abundance distributions of a given star \citep[see for
instance][]{SilvesterSiKoWa2014b}. It is clear that current modelling not only
of atomic diffusion but of ZDM as well must have their respective limitations.
In this paper, we will discuss only the limitations in theoretical modelling. An
improved grid of equilibrium abundance distributions of 16 chemical elements
that result from the standard theoretical diffusion model is presented in
Sec.\,\ref{Abequilb}. We then discuss in Sec.\,\ref{depsimp} extensions to the
standard model that may be encountered in real stars, providing possible
explanations for substantial deviations from observational predictions based on
the standard model.

\section{Equilibrium abundance stratifications}
\label{Abequilb}

Theoretical atmospheric models which take into account chemical stratifications
due to atomic diffusion and which are based on extensive calculations of
radiative accelerations have first been proposed by \citet{AlecianAlSt2007g} and
by \citet{LeBlancLeMoHuetal2009l}. In a later article, \citet{AlecianAlSt2010}
established bi-dimensional distributions of elements in magnetic atmospheres,
assuming a dipolar geometry for the magnetic fields; variations in (vertical)
abundance stratifications along the magnetic meridian -- from the magnetic pole
to the equator -- were presented for 16 metals. The results discussed in the
above-mentioned articles are all based on so called \emph{equilibrium}
stratifications representing the abundance stratifications necessary to have
zero diffusion velocity for each element (remember that, at least in optically
thick layers, the diffusion velocity largely depends on the local abundance of
the element). This is approximately equivalent to the module of the radiative
acceleration vector being equal to the module of the vector of gravity (the
vectors are of opposite sign\footnote{Theoretical considerations show that these
vectors could point in the same direction only in a very particular and scarce
case of a radiative acceleration dominated by some free-free transitions
\citep{MassacrierMa2005}.}).

Our criterion for equilibrium thus consists in achieving an effective total
acceleration \citep[as defined by Eq.\,15 of][]{AlecianAlSt2006i} $g_{\rm
tot}^{\rm eff} = 0$ for each element by means of an iterative modification of
element stratifications. Let us emphasise the fact that an \emph{equilibrium}
calculation yields the maximum abundance of an element that can be supported by
the radiation field in a given atmospheric layer. Therefore the equilibrium
stratification cannot be considered to correspond to the one that would be
obtained through the regular physical process of stratification build-up which
obeys the time-dependent continuity equation. Time-dependent solutions vs.
equilibrium stratifications will be discussed in a forthcoming paper.

\subsection{New calculations for a $T_{\rm eff}=10\,000$\,K atmosphere}
\label{newcalc}

The results of \citet{AlecianAlSt2010} have been obtained with fixed atmospheric
models based on solar abundances of the chemical elements. More recently,
\cite{StiftStAl2012} developed a modified {\sc CaratStrat} code which ensures
self-consistency in the calculation of equilibrium stratifications; the final
vertical abundance distributions are now consistent with the stratified
atmospheric structure computed with {\sc Atlas12} (\citealt{Kurucz2005},
\citealt{Bischof2005}).
 
In Fig.\,\ref{fig:Fig_Layout_strat_eq} we present the equilibrium
stratifications for 16 elements (adopting a main sequence atmospheric model with
$T_{\rm eff}=10\,000$\,K), obtained with this new self-consistent version of
{\sc CaratStrat}. The red dashed curves show the equilibrium stratifications in
the presence of vertical magnetic lines of 20\,kG (hereinafter considered as the
magnetic pole of the stellar dipole), the red solid curves pertain to horizontal
magnetic lines of 10\,kG (corresponding to the magnetic equator of the same
dipole). The black curves display the results for intermediate magnetic angles
($60\,\degr$ and $80\,\degr$) for which we have used available calculations for
a grid of models (not restricted to a centred dipole geometry) with a field
strength of 10\,kG. For a dipole, the exact field strengths at these angles
would actually be 11.1\,kG and 10.1\,kG respectively; the field strength of
10\,kG for both $60\,\degr$ and $80\,\degr$ however is close enough to the exact
ones for our present purpose. These 4 curves will hopefully help the reader to
understand the predicted chemical stratifications as a function of the field
geometry in the atmosphere of a star with a strong magnetic field. Comparison
between the red dashed curves and the red solid ones reveals the abundance
differences between the magnetic pole and the magnetic equator in the case of a
dipolar magnetic field. Comparing the red curves to the black ones gives an
indication as to the expected abundance contrast over the stellar surface. The
arrows flag the layers for which convergence towards equilibrium has not been
fully achieved (our criterion is $\left|\log (g_{rad}/g)\right| > 0.1$). These
arrows point in the direction the curves should move to -- by about 0.3\,dex or
more -- in order to reach equilibrium, assuming that the behaviour of the
radiative accelerations corresponds to the optically thick case (which however
is not necessarily the case in optically thin layers). Generally, satisfactory
convergence has been obtained for all inclinations\footnote{In
Fig.\,\ref{fig:Fig_Layout_strat_eq}, Mg, Si, Cu, Hg have not perfectly converged
for $80\,\degr$ and $\log{\tau}<-3.0$\,. Notice that in our model, Mg and Ca are
never supported by the radiative acceleration in layers deeper than $\log\tau
\approx 1.0$, therefore, equilibrium cannot be reached for physical reasons.} of
magnetic lines less than $80\,\degr$, but it turns out especially difficult to
obtain convergence for $90\,\degr$. This explains why the arrows are only shown
for horizontal magnetic lines.

We have identified several numerical reasons why equilibrium is not attained:

(i) Optically thin layers can suffer from screening of the radiation field by
underlying layers. When the concentration of a given element becomes very strong
in layers at $\log{\tau} \approx -3.0$\, for instance, photons originating from
hotter deep layers are screened by the saturated lines of this element. Even if
they are re-emitted, they are not necessarily re-emitted at the same
temperature. This effect has well been identified in time-dependent diffusion
calculations \citep[see][]{AlecianAlStDo2011}, but ideally it should not affect
the search for equilibrium stratifications. However, our convergence algorithm
appears sensitive to it because each iteration is subdivided into two sub-steps:
in a first step an uniform increase of abundance is imposed on all layers,
allowing us to determine how the medium reacts; in a second step the new
abundance is determined by extra/interpolation. Screening may appear in the
first sub-step, and so it impacts on the second sub-step. Most of the cases
(arrows) shown in Fig.\,\ref{fig:Fig_Layout_strat_eq} are due this effect.

(ii) In some layers, diffusion coefficients of ionised particles across
horizontal magnetic lines become very small as the proton density decreases.
Since diffusion coefficients enter the expression of effective total
acceleration \citep[see Sec.\,5.1 of][]{AlecianAlSt2006i}, vanishing
coefficients become numerically very demanding on the convergence procedure
(especially in the optically thin case), either because of numerical
instabilities or because of an excessive increase in CPU time.

(iii) Another source of numerical instability consists in the difficulty
experienced by the {\sc Atlas12} module to deal with extremely strong abundance
gradients which cause temperature inversions, or with very high over-abundances
of metals. In some cases the equilibrium stratification may correspond to metal
abundances comparable to the abundance of hydrogen, which is clearly unsupported
by the code. For that reason we have put an upper limit of 9.5 (on a scale where
[H] = 12.00) to metal abundances and we forbid a negative photon flux.

\subsection{Results}
\label{results}

\begin{figure*}
\centering
\includegraphics[width=15cm]{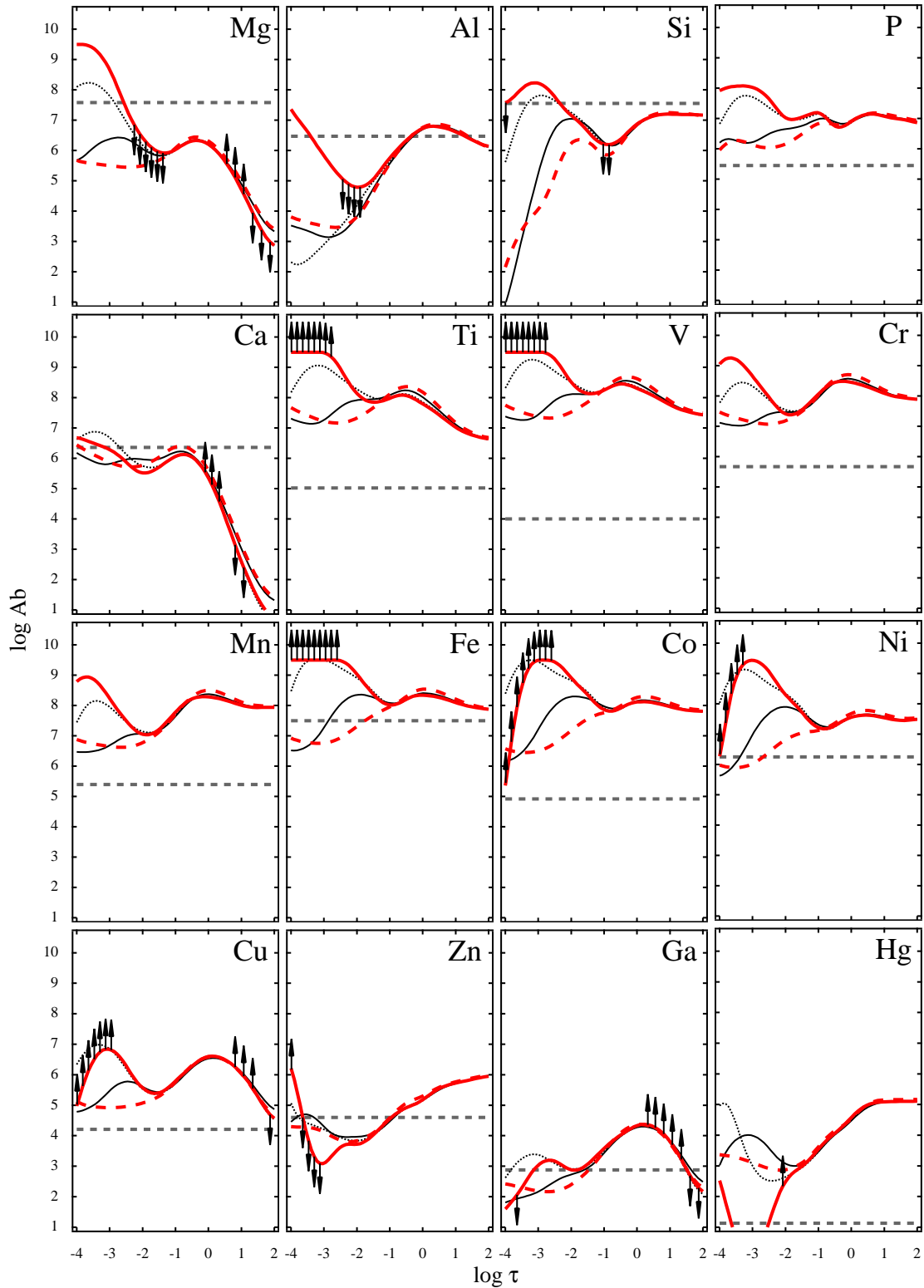}
\caption{
Logarithm of equilibrium abundances (given relative to hydrogen with
$\log\,H=12$) vs. logarithm of the optical depth (at 5000\,{\AA}) for 16
elements. The model atmosphere is characterised by $T_{\rm eff} = 10\,000$\,K
and $\log g = 4.0$. The horizontal grey heavy dashed line depicts the solar
abundance (uniform), the red heavy dashed line shows the abundance
stratification at the magnetic pole (20\,kG, $0\,\degr$), the red heavy solid
line pertains to the magnetic equator (10\,kG, $90\,\degr$). The other curves
correspond to intermediate magnetic parameters: the black solid line to 10\,kG,
$60\,\degr$, the black dotted one to 10\,kG, $80\,\degr$. The arrows attached to
the $90\,\degr$ curves indicate the layers where equilibrium has not been
reached (see text).
}
\label{fig:Fig_Layout_strat_eq}
\end{figure*}

Despite the above mentioned convergence problems, we consider the results shown
in Fig.\,\ref{fig:Fig_Layout_strat_eq} entirely adequate for a discussion of the
main abundance trends predicted by present-day theoretical models of magnetic
ApBp stars. Stratifications have been derived for the same elements as those
considered in \citet{AlecianAlSt2010}. In contrast to this earlier work, the
atmospheric model with $T_{\rm eff}=10\,000$\,K and $\log{g}=4.0$ has been
updated at every iteration step according to the abundance changes (see details
in \citealt{StiftStAl2012}).

The following discussion of the distributions of the different chemical elements
assumes that equilibrium stratifications can actually be reached in real
atmospheres and that NLTE conditions do not have a large effect on radiative
accelerations (see a discussion of the limits of these assumptions in
Sec.\,\ref{depsimp}). Please note that our results are rather sensitive to the
effective temperature and to the gravity (as a balance term against radiative
acceleration), but also to a lesser degree to the strength of the magnetic field
(due to Zeeman desaturation). They cannot therefore be generalised to all ApBp
stars and must be used with extreme caution when confronting them with
observations. Also keep in mind that a simple dipolar geometry is assumed
throughout even though detailed magnetic mapping of ApBp stars appear to reveal
much more complex magnetic geometries. Hereinafter, we shall use the word
\emph{stratification} for vertical abundance inhomogeneities, the word
\emph{distribution} for horizontal ones. We use the word \emph{cloud}  for both
stratification and distribution when abundance inhomogeneities reach a well
discernible contrast.

\bigskip\noindent {\bf{Mg}}: Magnesium is deficient everywhere in the atmosphere
except for $\log\tau < -2$ when magnetic lines are close to horizontal
(inclination of lines $> 80\,\degr$). High altitude clouds may be expected with
high contrast (+2\,dex) in zones with horizontal field lines. These clouds will
manifest themselves as a ring/belt around the star (along the magnetic equator).
Notice that  \ion{Mg}{3} (noble gas configuration with weak radiative
acceleration) is the dominant ion in layers deeper than $\log\tau \approx 0$, so
Mg is not supported in the deep atmosphere.

\bigskip\noindent {\bf{Al}}: Aluminium should form a cloud-like narrow ring/belt
around the star (along the magnetic equator) located at $\log\tau \approx -3$. A
horizontal magnetic field prevents Al ions from sinking, so the cloud is
supported by the radiative acceleration of the neutral state (the same effect as
invoked for silicon, see below). Al is expected to be strongly stratified
because it is not supported by the radiation field around $\log\tau \approx
-1.5$ where it becomes strongly deficient (the radiative acceleration is very
weak for \ion{Al}{2}).

\bigskip\noindent {\bf{Si}}: Silicon at solar abundance is not supported by the
radiation field except when the magnetic field is almost horizontal. This strong
dependence on the inclination of the magnetic field lines is consistent with
previous predictions \citep{VauclairVaHaPe1979,AlecianAlVa1981}. However, the
total radiative acceleration on Si is not large enough in our computations to
explain the over-abundances observed in ApSi stars. This problem has been
pointed out in the past by \citet{MichaudMi1970y} who proposed that
auto-ionisation lines could play an important role. \citet{AlecianAlVa1981}
evaluated the importance of the \ion{Si}{2} auto-ionisation lines but the atomic
data available in the 1980's \citep{ArtruArPrJaetal1981} did not allow to
conclude positively on their role. The question still remains open since {\sc
CaratStrat} does not consider these particular transitions.

\bigskip\noindent {\bf{P}}: Phosphorus is well supported by the radiation field
throughout the atmosphere. Its equilibrium abundance lies about 1\,dex above the
solar abundance with a rather smooth abundance stratification and a fairly
uniform distribution. It does not appear to be very sensitive to the magnetic
field. This is due to the rather high ionisation potential of \ion{P}{1}
(10.49\,eV) which leads (for $-4.0 < \log\tau < 0.0$) to a significantly larger
relative population of the neutral state than in the 15 other elements of our
sample (except Hg). Let us point out that the equilibrium is hardly affected at
all by the magnetic field because of the field-independence of the diffusion
coefficient of the neutral state. This effect is further enhanced by the fact
that the \ion{P}{2} atomic lines are saturated (in contrast to those of
\ion{Hg}{2}), thus the weight of \ion{P}{1} in the total radiative acceleration
is large.

\bigskip\noindent {\bf{Ca}}: Calcium is not strongly supported by the radiation
field because the dominant ion is \ion{Ca}{3} in all layers of the atmosphere.
\ion{Ca}{3} exhibits a noble gas configuration with ensuing very small radiative
acceleration. As a consequence, Ca has to be under-abundant everywhere to
satisfy the equilibrium condition. Ca is atypical because for the other elements
we find that the dominant ion (in layers $-4.0<\log\tau<0.0$) is generally the
first ionisation state. The equilibrium stratification of Ca was calculated for
the same effective temperature as the one adopted by
\citet{BorsenbergerBoPrMi1981} who took NLTE effects into account, but no
magnetic field. Our results are quite close to theirs, regarding the position of
the two abundance maxima. However, since their model atmosphere was computed
assuming homogeneous solar abundances for all elements, we cannot compare the
details of the respective Ca stratifications.

\bigskip\noindent {\bf{Ti}} and {\bf{V}}: Titanium and vanadium display very
similar equilibrium stratifications/distributions (the respective ionisation
potentials of their first three ionisation states differ little). These elements
should be over-abundant throughout the atmosphere with the largest enhancement
taking the form of a moderately narrow ($\approx \pm10\degr$) ring/belt around
the star (along the magnetic equator) above $\log\tau \approx -2$.

\bigskip\noindent {\bf{Cr}} and {\bf{Mn}}: Chromium and manganese are also well
supported by the radiation field, since an over-abundance of about 2\,dex is
needed before radiative acceleration balances gravity, even near the magnetic
pole. We note that Cr and Mn could accumulate by up to 4\,dex with respect to
the solar value above $\log\tau \approx -2$ in a rather narrow belt/ring along
the magnetic equator. Let us point out that these values are consistent with
early calculations for Mn in the non-magnetic case \citep{AlecianAlMi1981} and
with observations (see Fig.\,9 of \citealt{SmithSmDw1993r}).

\bigskip\noindent {\bf{Fe}}: A high abundance of iron is well supported over the
entire stellar surface. However, because of the high solar iron abundance, the
lines are already highly saturated even before any stratification build-up due
to atomic diffusion. Therefore, saturation effects on the radiative acceleration
are fully effective at the very beginning of the modelling procedure and
equilibrium is reached after only a few iterations for $\log\tau > -1.0$. This
however is not the case in optically thin layers where the magnetic field may
lead to a large ($ > \pm10\degr$) over-abundant ring/belt around the magnetic
equator. In the polar region, Fe can be under-abundant in upper atmospheric
layers. Notice that inside the ring/belt, the abundance of iron attains the
upper limit of [Fe] = 9.5 (with [H] = 12.00) we have fixed.

\bigskip\noindent {\bf{Co}} and {\bf{Ni}}: Cobalt and nickel are less abundant
in the upper layers of magnetic polar regions than in deep atmospheric layers
($\log\tau > -1.0$) where they could be uniformly strongly over-abundant.
However, far from the polar regions (field angle $ > 60\degr$), Co and Ni
accumulate and should form a large belt above $\log\tau \approx -2.0$ ($ >
\pm30\degr$) around the equator.

\bigskip\noindent {\bf{Cu}}: Copper shows a strong enhancement of its abundance 
(by about 2.5\,dex) around $\log\tau = 0$, uniformly distributed over the 
stellar surface and not depending on the magnetic field. However, a well 
contrasted equatorial ring/belt (3\,dex) may form above $\log\tau \approx -2.5$.

\bigskip\noindent {\bf{Zn}}: Zinc is over-abundant uniformly in layers deeper
than $\log\tau = 0$, and under-abundant uniformly in the interval
 $-2.5 < \log\tau < -0.5$. Far from the magnetic pole (field angle $> 60\degr$), 
Zn forms a belt around the star. However, this accumulation occurs in
high layers, just above the uniformly under-abundant region, so it is difficult 
to know what will dominate in the observed spectra (over- or under-abundance) 
and how this could possibly be diagnosed.

\bigskip\noindent {\bf{Ga}}: Gallium is uniformly over-abundant ($\approx
+1.5$\,dex) in layers $-1.0 < \log\tau < +0.1$, and slightly under-abundant
($\approx -1.0$\,dex) above $\log\tau= -2.0$. The abundance seems to be
marginally dependent on the magnetic field for layers deeper than $\log\tau=
-3.0$. These results are consistent with the study of the Ga case in upper
main-sequence chemically peculiar (CP) stars by \citet{AlecianAlAr1987} (with
and without a magnetic field). Notice that these authors predicted Ga
over-abundances to increase with effective temperature in the range 
$10\,000\rm{K} < T_{\rm{eff}} < 15\,000\rm{K}$.

\bigskip\noindent {\bf{Hg}}: Mercury is a very special element in the study of
CP stars since it is found to be highly over-abundant in HgMn stars, but not
especially enhanced in spectra of magnetic ApBp stars. In the present
calculations, the equilibrium stratification of Hg exhibits a uniform
over-abundance (larger than $\approx +3$\,dex) in layers deeper than $\log\tau=
-1.0$, and a kind of hole around $\log\tau= -3.0$. This hole is much more
conspicuous near the magnetic equator. A cloud of Hg seems to form at the
magnetic equator above $\log\tau= -4.0$. As far as Hg is concerned, the
equilibrium stratification approach is unsatisfactory when it comes to an
explanation of the dichotomy between magnetic and non-magnetic stars, despite
the fact that a hole develops around $\log\tau= -3.0$ in the presence of
horizontal magnetic field lines. One should keep in mind the detailed study of
Hg radiative accelerations by \citet{ProffittPrBrLeetal1999r} which has shown
that Hg is strongly affected by non-LTE effects. Therefore, Hg should be
considered in a time-dependent diffusion and possibly non-LTE framework.

\section{Beyond the simplest theoretical model}
\label{depsimp}
The results discussed in previous section should be considered as an attempt 
to establish a kind of reference frame for abundance stratifications in a
$T_{\rm{eff}}=10\,000$ K, $\log g = 4.0$ star with a dipolar magnetic field.
The stratifications represent equilibrium values which are obtained through 
an iterative procedure. Since atomic diffusion is a process very sensitive 
to stellar properties, but also to any kind of perturbation, real stars 
could easily not conform to this simple model. In the following we shall 
shortly discuss several complications which can lead to deviations from the
abundance stratifications obtained with the simple model.

\subsection{NLTE effects}
\label{nlte}

The role of NLTE effects has a particular status among the processes we consider
in this section. In contrast to the processes enumerated below, NLTE is not just
another parameter to add to the model or to neglect. We know that above a
certain depth in the atmosphere -- which depends on the element -- NLTE effects
will be important enough to invalidate the results computed with the LTE based
{\sc{CaratStrat}} code. So far, for 3 of the elements discussed in
Sec.\,\ref{results} there exist studies dealing with NLTE effects on their
radiative accelerations. \citet{BorsenbergerBoPrMi1981} looked at Ca,
\citet{AlecianAlMi1981} at Mn. The paper on Hg by
\citet{ProffittPrBrLeetal1999r} constitutes the most detailed modelling attempt
ever of Hg radiative accelerations. It has been established that for the HgMn
star $\chi$\,Lupi, the observed abundance of mercury cannot be supported by the
radiation field if NLTE effects are properly taken into account. Note that their
calculations were done assuming a homogeneous overabundance of Hg. In the case
of Ca and Hg, NLTE calculations have revealed that radiative accelerations
within the framework of LTE are strongly overestimated in higher atmospheric
layers. Therefore, one should consider calculations of LTE based radiative
accelerations in upper atmospheric layers with the utmost caution. It is for
this reason that the results shown in Fig.\,\ref{fig:Fig_Layout_strat_eq} are
plotted only for layers with $\log\tau > -4.0$.

\subsection{Decentred dipole}
\label{decenter}
The hypothesis of a magnetic geometry characterised by a centred dipole implies
that the star must have symmetrical magnetic hemispheres. If however the dipole
is displaced from the centre of the star \citep{LandstreetLa1970r}, the magnetic
equator will no longer delimit 2 symmetrical hemispheres. The magnetic field
could be much stronger at one pole (leading to significantly larger
amplifications of radiative accelerations) than at the other pole. Asymmetric
abundance distributions are often found in ApBp stars (see for instance,
\citealt{SilvesterSiKoWa2014b}). If diffusion were affected solely by the
magnetic field, such an asymmetry could only be explained  by the dipole being
off-centre or by more complex magnetic geometries.

\subsection{Multipolar fields}
\label{multi}
Although it is generally agreed that the dominant geometry of fossil magnetic 
fields is dipolar, this does not exclude the existence of quadrupolar or 
octupolar components \citep{MichaudMiChMe1981}. Thus the field lines will
not strictly follow a dipolar axial symmetry. Since the diffusion velocity is 
very sensitive to small variations in the angle between field line and stellar 
surface when the field is close to horizontal, quadrupolar and octupolar 
components can be expected to affect abundance distributions such as to make 
them patchier than in a strictly dipolar situation, but also to create
warped rings about the magnetic equator.

\subsection{Anisotropic wind}
\label{wind}
The mass loss velocity, assuming conservation of mass flux inside the star, is
found to vary with depth approximately as the inverse of mass density; the
diffusion coefficient for a given ion also varies more or less the same way. For
small mass loss rates we can neglect the wind velocity in the continuity
equation, for large mass loss rates the diffusion becomes negligible, but in
between, for an adequate mass loss rate, both velocities can compete.
Competition of atomic diffusion with a stellar wind (or with mass loss) has
first been considered by \citet{VauclairVa1975u} who invoked it to explain
He-rich stars. \citet{MichaudMiTaChetal1983q} and \citet{AlecianAl1996} included
mass loss in the study of Am stars. Evolutionary models which take into account
atomic diffusion, computed with the Montreal code \citep{VickViMiRietal2010}
include mass loss among the standard competitors to atomic diffusion.
\citet{BabelBa1992r} proposed a magnetically confined wind for 53\,Cam, because
he found that with atomic diffusion alone it was not possible to explain the
observed abundance distributions in their entirety. The assumption of an
anisotropic mass loss of about $3 \times {10^{-15}}$\Mloss close to the magnetic
pole significantly improved the consistency of the model with the observations
available at that time.

Our results shown in Fig.\,\ref{fig:Fig_Layout_strat_eq} are parameter free, and 
thus assume zero mass loss velocity. It is clear that individual stars need to 
be modelled assuming non-zero mass loss velocity, in particular magnetic stars 
with confined anisotropic wind near the magnetic poles. Such an approach could 
certainly strongly modify the polar abundances shown in 
Fig.\,\ref{fig:Fig_Layout_strat_eq}.

\subsection{Mixing}
\label{mixing}

In the modelling of ApBp stars (with and without magnetic fields), it is always
assumed that the atmosphere is completely stable (note that in AmFm stars the
atmospheres are convective, and so models including diffusion are restricted to
internal layers). We will not develop here the arguments in favour of this
universally accepted assumption; it can be justified from theoretical
considerations, but empirical support comes from observational evidence -- which
has accumulated over recent years -- for abundance stratifications. Still, one
cannot completely rule out that weak mixing (including thermohaline convection,
see \citealt{VauclairVa2004}) may take place, perhaps only over a few layers.

It is difficult to assess what happens in the atmosphere of a magnetic ApBp
star with strong vertical stratifications depending on the angle between the 
field vector and the surface normal. If the local atmospheres were isolated,
the horizontal element distributions would differ from layer to layer, and the 
local temperature structure would vary accordingly; electron density and
gas pressure would not be the same for a given depth. Basic physics of course
tell us that pressure equilibrium between the local atmospheres will be 
established on very short time-scales. Needless to say that nobody yet has 
pointed out this problem, even less addressed the question of how the stellar 
atmosphere manages to stay in equilibrium horizontally and vertically in the
presence of such chemical inhomogeneities. Does this give rise to some large 
scale horizontal circulation (e.g. similar to a vortex), resulting in
horizontal mixing?

\subsection{Other processes}
\label{other}

Keeping in mind that real stars will behave in a much more complex way than 
the various models, and knowing that atomic diffusion is easily perturbed, 
we have to realise that the list of physical processes which can modify the 
simple picture proposed in Fig.\,\ref{fig:Fig_Layout_strat_eq} is certainly 
larger than the enumeration given above. 

Processes like accretion of diffuse matter or the infall of solid bodies need 
to be looked at. Accretion of gas/dust may be the preferred scenario to atomic
diffusion for $\lambda$\,Boo type stars \citep[see][]{TurcotteTuCh1993y}, 
but not so for ApBp stars. Still, infall of solid bodies onto ApBp stars 
(including HgMn stars) may occasionally lead to a modification of the surface 
element distributions and abundances \citep{Cowley1977}. The relaxation time 
-- by atomic diffusion -- of such events remains to be estimated; it will 
involve hydrodynamics and depend on diffusion times (which are different for 
each element).

Tidal effects may also be considered, especially for HgMn stars which are 
slightly more often found in double-lined spectroscopic binary systems than 
normal stars, frequently with eccentric orbits \citep{SmithSm1996}. Even if 
tidal effects do not appear to lead to synchronisation of rotation for these 
binaries, they could possibly affect the stability of the atmosphere to a 
sufficient degree to cause detectable effects.

Among main-sequence CP stars, pulsations are found at least in roAp stars, and
possibly in HgMn stars which partly share the region of slowly pulsating B type
(SPB) stars in the HR-diagram \citep{AlecianAlGeAuetal2009,MorelMoBrAuetal2014};
for roAp stars, pulsations seem to coexist with vertical abundance
stratifications \citep[see for instance][]{FreyhammerFrKuEletal2009}. It is not
yet known how these pulsations interact with the atmospheric structure, and
whether or not they induce weak mixing.

First however, prior to the exploration and incorporation into our models of any
of the processes listed above, we have to keep in mind one of the main aspects
of atomic diffusion which is not dealt with in this article: the time dependent
nature of the build-up of chemical stratifications. This will be discussed in a
forthcoming paper.

\section{Conclusions}
\label{conclusion}

We have presented and discussed a comprehensive set of equilibrium 
stratifications of 16 chemical elements in the atmosphere of a magnetic 
star with $T_{\rm eff} = 10\,000$\,K and $\log g = 4.0$; field strengths of 
10 and 20\,kG were considered, with field lines inclined by $90\,\degr$,
$80\,\degr$, $60\,\degr$ and $0\,\degr$ towards the surface normal. It emerges 
from these simulations that in the majority of cases, large over-abundances 
are found in higher atmospheric layers when the fields are almost horizontal. 
Given the sensitivity of the stratifications to the field angle, we expect very
large over-abundances to show up mainly along and near the magnetic equator in 
upper atmospheric layers -- some patchy structure may be due to slight
deviations of the field direction from the surface normal -- more moderate ones
around the magnetic poles. We have emphasised the fact that these results 
strongly depend on the fundamental parameters of a star; they should thus not 
be extended indiscriminately to all kinds and classes of CP stars with their
considerable range in effective temperature and gravity. 

It must be admitted that at present this kind of modelling based on equilibrium
calculations is far too simple. Besides the fact that equilibrium calculations
do not incorporate the continuity equation, we have listed a number of physical
processes (such as NLTE, multipolar fields, anisotropic wind, etc.) which can
strongly affect the build-up of abundance stratifications and which should be
introduced into numerical codes dealing with atomic diffusion in atmospheres
(including {\sc{CaratStrat}}). The increase in the number of free parameters
constitutes of course a drawback.

We think that time-dependent approach should be the most realistic approach, but
it still needs further development (work is in progress). Notice also that
time-dependent calculations are extremely sensitive both to boundary and initial
conditions. Besides possible physical instabilities in optically thin cases
\citep{AlecianAlStDo2011}, these calculations are faced with numerical stability
problems, viz. for layers where the magnetic field is horizontal. In contrast,
whereas the modelling of equilibrium stratifications has attained a state of
superior maturity, it can hardly ever be applied to a given real, observed
atmosphere (except perhaps in places where the field is almost horizontal)
because by definition it only gives the maximum abundance which can be supported
by radiative acceleration. Still, it offers a reference frame within which
general statistical abundance trends found in CP stars all along the
main-sequence can be analysed and interpreted.

Concerning the apparent disagreement between theoretical predictions and
recently published empirical maps, we think that it will be important to check
the ability of ZDM algorithms to detect complex abundance structures, in
particular those predicted by diffusion theory for simple configurations, such
as for instance small spots and thin, possibly warped rings. It will also be
necessary to verify the effects of inhomogeneous vertical and horizontal,
field-dependent abundance distributions (including self-consistent models of
atmospheres) on the inversion procedure. \textbf{Theoretical modelling on its side will have to consider more realistic magnetic structures such as those derived from observations.}

Despite appearances, the outlook is by no means bleak. Numerical modelling that
includes atomic diffusion has made tremendous progress over recent years,
establishing -- among others -- incontrovertibly the sensitivity to the
inclination of the magnetic field  lines in the build-up of vertical abundance
structure. Long before the sophisticated calculations presently possible,
equilibrium stratifications have predicted statistical trends in the abundance
anomalies observed in CP stars, as for instance the dependence of the maximum Mn
overabundance on the effective temperature in HgMn stars (confirmed by
\citealt{SmithSmDw1993r}). Time-dependent diffusion calculations
\citep{AlecianAlStDo2011} have unveiled the complex behaviour of the build-up of
abundance stratifications and have provided new insight into the physics of this
process.

\section*{Acknowledgements}

Most of the numerical results discussed in this paper have been obtained using
the numerical code {\sc{CaratStrat}} developed in collaboration and written
mostly by M.~J. Stift. Thanks also to him for constructive discussions in the
framework of our longstanding collaboration. This work has been supported by the
Programme National de Physique Stellaire (PNPS) of CNRS/INSU, and by 
Observatoire de Paris-Meudon in the framework of \emph{Actions F\'ed\'eratrices
Etoiles}. This work was partly performed using HPC resources from GENCI-CINES
(grants c2014045021, c2015045021). Thanks go to AdaCore for providing the GNAT
GPL Edition of its Ada2005 compiler.

\bsp
\bibliographystyle{mn2e}
\bibliography{diff-structure}

\begin{thebibliography}{}

\bibitem[\protect\citeauthoryear{{Alecian}}{{Alecian}}{1996}]{AlecianAl1996}
{Alecian} G.,  1996, A\&A, 310, 872

\bibitem[\protect\citeauthoryear{{Alecian} \& {Artru}}{{Alecian} \&
  {Artru}}{1987}]{AlecianAlAr1987}
{Alecian} G.,  {Artru} M.,  1987, A\&A, 186, 223

\bibitem[\protect\citeauthoryear{{Alecian}, {Gebran}, {Auvergne}, {Richard},
  {Samadi}, {Weiss} \& {Baglin}}{{Alecian}
  et~al.}{2009}]{AlecianAlGeAuetal2009}
{Alecian} G.,  {Gebran} M.,  {Auvergne} M.,  {Richard} O.,  {Samadi} R.,
  {Weiss} W.~W.,    {Baglin} A.,  2009, A\&A, 506, 69

\bibitem[\protect\citeauthoryear{{Alecian} \& {Michaud}}{{Alecian} \&
  {Michaud}}{1981}]{AlecianAlMi1981}
{Alecian} G.,  {Michaud} G.,  1981, ApJ, 245, 226

\bibitem[\protect\citeauthoryear{{Alecian} \& {Stift}}{{Alecian} \&
  {Stift}}{2006}]{AlecianAlSt2006i}
{Alecian} G.,  {Stift} M.~J.,  2006, A\&A, 454, 571

\bibitem[\protect\citeauthoryear{{Alecian} \& {Stift}}{{Alecian} \&
  {Stift}}{2007}]{AlecianAlSt2007g}
{Alecian} G.,  {Stift} M.~J.,  2007, A\&A, 475, 659

\bibitem[\protect\citeauthoryear{{Alecian} \& {Stift}}{{Alecian} \&
  {Stift}}{2010}]{AlecianAlSt2010}
{Alecian} G.,  {Stift} M.~J.,  2010, A\&A, 516, A53+

\bibitem[\protect\citeauthoryear{{Alecian}, {Stift} \& {Dorfi}}{{Alecian}
  et~al.}{2011}]{AlecianAlStDo2011}
{Alecian} G.,  {Stift} M.~J.,    {Dorfi} E.~A.,  2011, MNRAS, 418, 986

\bibitem[\protect\citeauthoryear{{Alecian} \& {Vauclair}}{{Alecian} \&
  {Vauclair}}{1981}]{AlecianAlVa1981}
{Alecian} G.,  {Vauclair} S.,  1981, A\&A, 101, 16

\bibitem[\protect\citeauthoryear{{Artru}, {Praderie}, {Jamar} \&
  {Petrini}}{{Artru} et~al.}{1981}]{ArtruArPrJaetal1981}
{Artru} M.~C.,  {Praderie} F.,  {Jamar} C.,    {Petrini} D.,  1981, A\&A, 96,
  380

\bibitem[\protect\citeauthoryear{{Babcock}}{{Babcock}}{1949a}]{Babcock1949}
{Babcock} H.~W.,  1949a, The Observatory, 69, 191

\bibitem[\protect\citeauthoryear{{Babcock}}{{Babcock}}{1949b}]{Babcock_PASP_1949}
{Babcock} H.~W.,  1949b, PASP, 61, 226

\bibitem[\protect\citeauthoryear{{Babcock}}{{Babcock}}{1958}]{Babcock1958}
{Babcock} H.~W.,  1958, ApJ, 128, 228

\bibitem[\protect\citeauthoryear{{Babcock} \& {Burd}}{{Babcock} \&
  {Burd}}{1952}]{BabcockBurd1952}
{Babcock} H.~W.,  {Burd} S.,  1952, ApJ, 116, 8

\bibitem[\protect\citeauthoryear{{Babel}}{{Babel}}{1992}]{BabelBa1992r}
{Babel} J.,  1992, A\&A, 258, 449

\bibitem[\protect\citeauthoryear{{Bailey}, {Landstreet} \& {Bagnulo}}{{Bailey}
  et~al.}{2014}]{BaileyBaLaBa2014l}
{Bailey} J.~D.,  {Landstreet} J.~D.,    {Bagnulo} S.,  2014, A\&A, 561, A147

\bibitem[\protect\citeauthoryear{{Bischof}}{{Bischof}}{2005}]{Bischof2005}
{Bischof} K.~M.,  2005, Memorie della Societa Astronomica Italiana Supplementi,
  8, 64

\bibitem[\protect\citeauthoryear{{Borsenberger}, {Praderie} \&
  {Michaud}}{{Borsenberger} et~al.}{1981}]{BorsenbergerBoPrMi1981}
{Borsenberger} J.,  {Praderie} F.,    {Michaud} G.,  1981, ApJ, 243, 533

\bibitem[\protect\citeauthoryear{{Cowley}}{{Cowley}}{1977}]{Cowley1977}
{Cowley} C.~R.,  1977, Ap\&SS, 51, 349

\bibitem[\protect\citeauthoryear{{Deutsch}}{{Deutsch}}{1956}]{DeutschDe1956}
{Deutsch} A.~J.,  1956, PASP, 68, 92

\bibitem[\protect\citeauthoryear{{Freyhammer}, {Kurtz}, {Elkin}, {Mathys},
  {Savanov}, {Zima}, {Shibahashi} \& {Sekiguchi}}{{Freyhammer}
  et~al.}{2009}]{FreyhammerFrKuEletal2009}
{Freyhammer} L.~M.,  {Kurtz} D.~W.,  {Elkin} V.~G.,  {Mathys} G.,  {Savanov}
  I.,  {Zima} W.,  {Shibahashi} H.,    {Sekiguchi} K.,  2009, MNRAS, 396, 325

\bibitem[\protect\citeauthoryear{{Khokhlova}}{{Khokhlova}}{1994}]{KhokhlovaKh1994}
{Khokhlova} V.~L.,  1994, in {Zverko} J.,  {Ziznovsky} J.,  eds, Chemically
  Peculiar and Magnetic Stars {On the Comparison of Different Codes for CP
  Stars Surface Mapping}.
p.~60

\bibitem[\protect\citeauthoryear{{Kurucz}}{{Kurucz}}{2005}]{Kurucz2005}
{Kurucz} R.~L.,  2005, Memorie della Societa Astronomica Italiana Supplementi,
  8, 14

\bibitem[\protect\citeauthoryear{{Landstreet}}{{Landstreet}}{1970}]{LandstreetLa1970r}
{Landstreet} J.~D.,  1970, ApJ, 159, 1001

\bibitem[\protect\citeauthoryear{{LeBlanc}, {Monin}, {Hui-Bon-Hoa} \&
  {Hauschildt}}{{LeBlanc} et~al.}{2009}]{LeBlancLeMoHuetal2009l}
{LeBlanc} F.,  {Monin} D.,  {Hui-Bon-Hoa} A.,    {Hauschildt} P.~H.,  2009,
  A\&A, 495, 937

\bibitem[\protect\citeauthoryear{{Massacrier}}{{Massacrier}}{2005}]{MassacrierMa2005}
{Massacrier} G.,  2005, in {Alecian} G.,  {Richard} O.,   {Vauclair} S.,  eds,
  EAS Publications Series Vol.~17 of EAS Publications Series, {Radiative
  accelerations through photoionization and bremsstrahlung; autoionization}.
pp 61--66

\bibitem[\protect\citeauthoryear{{Michaud}}{{Michaud}}{1970}]{MichaudMi1970y}
{Michaud} G.,  1970, ApJ, 160, 641

\bibitem[\protect\citeauthoryear{{Michaud}, {Charland} \&
  {Megessier}}{{Michaud} et~al.}{1981}]{MichaudMiChMe1981}
{Michaud} G.,  {Charland} Y.,    {Megessier} C.,  1981, A\&A, 103, 244

\bibitem[\protect\citeauthoryear{{Michaud}, {Tarasick}, {Charland} \&
  {Pelletier}}{{Michaud} et~al.}{1983}]{MichaudMiTaChetal1983q}
{Michaud} G.,  {Tarasick} D.,  {Charland} Y.,    {Pelletier} C.,  1983, ApJ,
  269, 239

\bibitem[\protect\citeauthoryear{{Morel}, {Briquet}, {Auvergne}, {Alecian},
  {Ghazaryan}, {Niemczura}, {Fossati}, {Lehmann}, {Hubrig}, {Ulusoy},
  {Damerdji}, {Rainer}, {Poretti}, {Borsa}, {Scardia} \& et. al.}{{Morel}
  et~al.}{2014}]{MorelMoBrAuetal2014}
{Morel} T.,  {Briquet} M.,  {Auvergne} M.,  {Alecian} G.,  {Ghazaryan} S.,
  {Niemczura} E.,  {Fossati} L.,  {Lehmann} H.,  {Hubrig} S.,  {Ulusoy} C.,
  {Damerdji} Y.,  {Rainer} M.,  {Poretti} E.,  {Borsa} F.,  {Scardia} M.,
  et. al. 2014, A\&A, 561, A35

\bibitem[\protect\citeauthoryear{{Proffitt}, {Brage}, {Leckrone}, {Wahlgren},
  {Brandt}, {Sansonetti}, {Reader} \& {Johansson}}{{Proffitt}
  et~al.}{1999}]{ProffittPrBrLeetal1999r}
{Proffitt} C.~R.,  {Brage} T.,  {Leckrone} D.~S.,  {Wahlgren} G.~M.,  {Brandt}
  J.~C.,  {Sansonetti} C.~J.,  {Reader} J.,    {Johansson} S.~G.,  1999, ApJ,
  512, 942

\bibitem[\protect\citeauthoryear{{Ryabchikova}}{{Ryabchikova}}{2008}]{Ryabchikova2008}
{Ryabchikova} T.,  2008, Contributions of the Astronomical Observatory Skalnate
  Pleso, 38, 257

\bibitem[\protect\citeauthoryear{{Silvester}, {Kochukhov} \&
  {Wade}}{{Silvester} et~al.}{2014}]{SilvesterSiKoWa2014b}
{Silvester} J.,  {Kochukhov} O.,    {Wade} G.~A.,  2014, MNRAS, 444, 1442

\bibitem[\protect\citeauthoryear{{Smith}}{{Smith}}{1996}]{SmithSm1996}
{Smith} K.~C.,  1996, Ap\&SS, 237, 77

\bibitem[\protect\citeauthoryear{{Smith} \& {Dworetsky}}{{Smith} \&
  {Dworetsky}}{1993}]{SmithSmDw1993r}
{Smith} K.~C.,  {Dworetsky} M.~M.,  1993, A\&A, 274, 335

\bibitem[\protect\citeauthoryear{{Stift} \& {Alecian}}{{Stift} \&
  {Alecian}}{2012}]{StiftStAl2012}
{Stift} M.~J.,  {Alecian} G.,  2012, MNRAS, 425, 2715

\bibitem[\protect\citeauthoryear{{Turcotte} \& {Charbonneau}}{{Turcotte} \&
  {Charbonneau}}{1993}]{TurcotteTuCh1993y}
{Turcotte} S.,  {Charbonneau} P.,  1993, ApJ, 413, 376

\bibitem[\protect\citeauthoryear{{Vauclair}}{{Vauclair}}{1975}]{VauclairVa1975u}
{Vauclair} S.,  1975, A\&A, 45, 233

\bibitem[\protect\citeauthoryear{{Vauclair}}{{Vauclair}}{2004}]{VauclairVa2004}
{Vauclair} S.,  2004, in {Zverko} J.,  {Ziznovsky} J.,  {Adelman} S.~J.,
  {Weiss} W.~W.,  eds, The A-Star Puzzle Vol.~224 of IAU Symposium,
  {Thermohaline convection and metallic fingers in polluted stars}.
pp 161--166

\bibitem[\protect\citeauthoryear{{Vauclair}, {Hardorp} \&
  {Peterson}}{{Vauclair} et~al.}{1979}]{VauclairVaHaPe1979}
{Vauclair} S.,  {Hardorp} J.,    {Peterson} D.~M.,  1979, ApJ, 227, 526

\bibitem[\protect\citeauthoryear{{Vick}, {Michaud}, {Richer} \&
  {Richard}}{{Vick} et~al.}{2010}]{VickViMiRietal2010}
{Vick} M.,  {Michaud} G.,  {Richer} J.,    {Richard} O.,  2010, A\&A, 521, A62

\bibitem[\protect\citeauthoryear{{Vogt}, {Penrod} \& {Hatzes}}{{Vogt}
  et~al.}{1987}]{VogtVoPeHa1987}
{Vogt} S.~S.,  {Penrod} G.~D.,    {Hatzes} A.~P.,  1987, ApJ, 321, 496

\end{thebibliography}

\end{document}